# THEOREMS TO DEMOSTRATE THE PRESENCE OF ANTIFERROMAGNETISM IN THE PERIODIC ANDERSON MODEL

Omamoke O. E. Enaroseha
Nigeria Atomic Energy Commission, Abuja, Nigeria
Center for Nuclear Energy Studies, Port – Harcourt, Nigeria
enarosehaomamoke@gmail.com

and
Godfrey E. Akpojotor
Theoretical and Computational Condensed Matter Physics, Physics Department, Delta State University, Abraka 33106, Nigeria
akpogea@deltastate.edu.ng

Anderson model is an important model in the theory of strongly correlated electron system. In this study, we explore the ground state of this model and the concept of electron correlation by bipartite lattice and prove rigorously theorems leading to the presence of spin singlet in the model. By using the results of Ueda et al (1992) and Tian (1994), we show theoretically that the ground state of the symmetric periodic Anderson model has a short – range order antiferromagnetism.

## 1.0. INTRODUCTION

The role of physical models in investigating complicated systems is one of the most significant strategies in Theoretical Physics. The point is that if the model is able to account for some of the well known properties of the system, then it is expected to account for the more complex properties. Now when one is interested in fundamental properties of the model, exploring the ground-state energy is a feasible starting point [1]. For example, one of us has shown that for any finite value of the physically realistic onsite Coulombic interaction, the ground state of the Hubbard model which is the simplest model for studying strongly correlated systems, is a spin singlet [2]. Thus to use the model to study complicated strongly correlated phenomena like superconductivity and ferromagnetism which require transition from the antiferromagnetic ordering to partial or full spin ordering, other interaction terms needs to be added [3-4]. This is the line of thinking that led to the formation of the Periodic Anderson Model (PAM) [5-7] often used to investigate the heavy fermion systems which are well-known strongly correlated electron systems that exhibit unusual thermodynamic, magnetic, and transport properties [8].

Take a finite lattice $\wedge$ with $N_\wedge$ lattice points. Neglecting the orbital degeneracy for the localized electrons, the Hamiltonian of the PAM can be written in the form [9]

$$H_\wedge = -t\sum_\sigma \sum_{\langle i,j \rangle}(C_{i\sigma}^+ C_{j\sigma} + C_{j\sigma}^+ C_{i\sigma}) + \sum_{i,\sigma}\varepsilon_d n_{i\sigma} + V\sum_{i\sigma}(C_{i\sigma}^+ d_{j\sigma} + d_{j\sigma}^+ C_{i\sigma}) + U\sum_{i\in\wedge} n_{i\uparrow}n_{i\downarrow} \quad (1)$$

where $c_{i\sigma}^+$ ($C_{i\sigma}$) is the creation (annihilation) of electrons, which creates (annihilates) an itinerant electron of spin σ at site i. Similarly, $d_{i\sigma}^+$ and $d_{i\sigma}$ are the fermion operators for the localized electrons. $n_{i\sigma} = d_{i\sigma}^+ d_{i\sigma}$ and $\langle i,j \rangle$ denotes a pair of lattice sites, $\varepsilon_d$ is a local potential, V is the hybridization matrix element and U > 0 represent the on-site coulomb interaction for the localized electrons.

Since its introduction, the PAM and its variants constitute an important research topic in theoretical condensed matter physics, particularly in the context of strongly correlated electron system. Most of the many body techniques commonly used in condensed matter physics can be learnt in this context. Also there are some theoretical tools and concepts which apply to this model only [10-13]. The aim of this paper is to provide a smooth introduction in the models and prove that the ground state of the symmetric PAM is a spin singlet.

## 2.0 METHODOLOGY AND MATHEMATICAL FORMULATION

Lattice $\wedge$ is called bipartite with respect to the Hamiltonian $H_\wedge$ if it can be divided into two sublattices *A* and *B*, such that hopping of electrons does not occur between sites in the same sublattices. For a bipartite lattice, the signs of parameters *t* and V are not important for the mathematical analysis of this Hamiltonian. They can be changed by a unitary transformation. For definiteness, we shall choose *t > 0* and *V > 0* in the following.

To begin with, we first write the Hamiltonian $H_\wedge$ of the symmetric PAM in a generalized Hubbard Hamiltonian. For simplicity, let us consider a specific bipartite lattice: the two-dimensional square lattice $\wedge$ with N = L² lattice points (the lattice constant is taken to be unity). Take two identical copies of this square lattice, $\wedge_1$ and $\wedge_2$. We make a doubly layered lattice $\tilde{\Lambda}$ by connecting the corresponding lattice points of $\wedge_1$ and $\wedge_2$ with bonds of length 1. Now, each point of $\tilde{\Lambda}$ is labelled by r = ( i, m) where *m* = 1 or 2. Obviously, $\tilde{\Lambda}$ has $2N_\wedge$ lattice points. Next, we introduce new fermion operators $f_{r\sigma}$ by

$$f_{r\sigma} = \begin{cases} C_{i\sigma} & \text{if } m=1 \\ d_{i\sigma} & \text{if } m=2 \end{cases} \quad (2)$$

If we assume in Eq. (1) that, when $\varepsilon_d = \varepsilon_f = -U/2$, the Hamiltonian of the PAM is called Symmetric. With the definitions of $\tilde{\Lambda}$ and $f_{ra}$, the Hamiltonian $H_\Lambda$ of the symmetric PAM can be rewritten as the Hamiltonian of a generalized Hubbard model on $\tilde{\Lambda}$ by ignoring an uninteresting constant $UH_\Lambda/4$.

$$H_{\tilde{\Lambda}} = -\sum_\sigma \sum_{\langle rh \rangle} t_{rh} \left( f^+_{r\sigma} f_{h\sigma} + f^+_{h\sigma} f_{r\sigma} \right) + U \sum_{r \in \Lambda_2} \left( n_{r\uparrow} - \frac{1}{2} \right)\left( n_{r\downarrow} - \frac{1}{2} \right) \quad (3)$$

where the new hopping constants $t_{rh}$ are defined by

$$t_{rh} = \begin{cases} t & \text{if r and h are nearest neighbors in } \Lambda_1 \\ -V & \text{if r and h are different layers and connected by bond} \\ 0 & \text{otherwise} \end{cases} \quad (4)$$

It is easy to see $\tilde{\Lambda}$ is still bipartite with respect to $H_\Lambda$. Furthermore, we observe that the electrons in the second layer (the *d*–electron layer) do not hop and have an on-site Coulomb repulsion, while the electrons in the first layer are itinerant without interaction. That causes a little technical inconvenience. For this reason, we introduce an auxiliary interaction term

$$H_1 = \varepsilon \sum_{r \in \Lambda_1} \left( n_{r\uparrow} - \frac{1}{2} \right)\left( n_{r\downarrow} - \frac{1}{2} \right) \quad (5)$$

to the Hamiltonian $H_\Lambda$ and study the ground state of the new Hamiltonian $\tilde{H}_{\tilde{\Lambda}} = H_{\tilde{\Lambda}} + H_1$. In the final step, we let $\varepsilon \to 0$.

The purposes of the proofs are to show the existence of a short-range antiferromagnetic *d*-electron spin correlation in the ground state $\Psi_0(\varepsilon, U, \tilde{\Lambda})$ of the positive-($\varepsilon$, *U*) symmetric PAM at half filling. However, reflection positivity in the spin space does not hold in this case. Fortunately, by the well-known unitary particle-hole transformation

$$\hat{U}_0 f_{r\uparrow} \hat{U}_0^{-1} = f_{r\uparrow}, \hat{U}_0 f_{r\downarrow} \hat{U}_0^{-1} = \varepsilon(r) f^+_{r\downarrow} \quad (6)$$

where $\varepsilon(r) = 1$ if r $\in$ A and $\varepsilon(r) = -1$ if r $\in$ B, the Hamiltonian of the same form, which has reflection positivity in the spin space. Obviously, the transformed Hamiltonian has the same spectrum. In the following, we shall first study the ground state $\Psi_0(-\varepsilon, -U, \tilde{\Lambda})$ of the Hamiltonian $\tilde{H}_{\tilde{\Lambda}}(-\varepsilon, -U)$ at half filling and then transform it back to the corresponding ground state $\Psi_0(\varepsilon, \tilde{\Lambda})$ of $H_\Lambda(\varepsilon, U)$ by the inverse of the particle-hole transformation.

We notice that $\tilde{H}_{\tilde{\Lambda}}$ commutes with $N_\uparrow = \sum f^+_{r\uparrow} f_{r\uparrow}$ and $N_\downarrow = \sum f^+_{r\downarrow} f_{r\downarrow}$ respectively. Consequently, the Hilbert space of $H_\Lambda$ can be divided into numerous subspaces. Each of them is characterized by a pair of integers ($N_\uparrow = n_1$, $N_\downarrow = n_2$). For this Hamiltonian, following the formalism of Ref [14], we proved the following theorem.

*Theorem.* For any given even integer $N$, the ground state $\Psi_0$ of $\tilde{H}_{\tilde{\Lambda}}$ is unique and has quantum numbers $n_1 = n_2 = {}^N/_2$. Furthermore, $\Psi_0$ can be written as

$$\tilde{\Psi}_0(\frac{N}{2},\frac{N}{2}) = \sum_{\alpha,\beta} W_{\alpha\beta} \Psi^\alpha_\uparrow \otimes \Psi^\beta_\downarrow \tag{7}$$

where $\Psi^\alpha_\uparrow$ is an orthogonal real basis for one species of ${}^N/_2$ spinless fermions. Take $W = W_{\alpha\beta}$ as a $C^{N/2}_{2N_\Lambda} X C^{N/2}_{2N_\Lambda}$ matrix; then $W$ is Hermitian and positive definite.

In Ref. [14], the uniqueness of $\tilde{\Psi}_0(\frac{N}{2},\frac{N}{2})$ is the consequence of the positive definiteness of $W$, which was proved under the condition $\varepsilon \neq 0$. Ueda *et al* [15] removed this condition by exploiting a special topology of the PAM in addition to the reflection positivity in the spin space. Consequently the ground state of $\tilde{H}_{\tilde{\Lambda}}$ is still non-degenerate even if $\varepsilon = 0$. With knowledge of this fact, we are able to make our proof simpler by introducing $H_I$ and employing Lieb's theorem, whose proof is more direct. The theorem of Ref. [15] guarantees that the ground state $\Psi_0(\varepsilon, U, \tilde{\Lambda}) \to \tilde{\Psi}_0(0, U, \tilde{\Lambda})$ as $\varepsilon \to 0$.

*Theorem 1.* Let $A_r \equiv f_{r\uparrow} f_{r\downarrow}$. Then, for any two distinct lattice points r and h, we have

$$\langle \Psi_0(\frac{N}{2},\frac{N}{2}) | A^+_r A_h | \Psi_0(\frac{N}{2},\frac{N}{2}) \rangle \geq 0 \tag{8}$$

*Proof.* Taking into account the form of Eq. (7), that is, $\Psi_0$ (N/2, N/2), we have

$$\langle \Psi_0 | A^+_r A_h | \Psi_0 \rangle \equiv \langle \Psi_0 | f^+_{r\downarrow} f^+_{r\uparrow} f_{h\uparrow} f_{h\downarrow} | \rangle = \langle \Psi_0 | (f^+_{r\uparrow} f_{h\uparrow}) (f^+_{\sum h\downarrow} f_{h\downarrow}) | \rangle$$

$$= \sum_{\tau\delta} \sum_{\alpha\beta} \overline{W}_{\tau\delta} W_{\alpha\beta} \langle \Psi^\tau_\uparrow | f^+_{r\uparrow} f_{h\uparrow} | \Psi^\alpha_\uparrow \rangle \langle \Psi^\delta_\downarrow | f^+_{r\downarrow} f_{h\downarrow} | \Psi^\beta_\downarrow \rangle$$

$$= \sum_{\tau\delta} \sum_{\alpha\beta} W^+_{\delta\tau} W_{\alpha\beta} \langle \Psi^\tau | f^+_r f_h | \Psi^\alpha \rangle \overline{\langle \Psi^\delta | f^+_r f_h | \Psi^\beta \rangle}$$

$$= Tr(WMW^+)$$

where $M \equiv f_r^+ f_h \langle \Psi_\delta | f_r^+ f_h | \Psi_\beta \rangle = \overline{\langle \Psi_\delta | f_r^+ f_h | \Psi_\beta \rangle}$ and $W^+ = W$ (9)

$$= \sum_{\tau\delta} \sum_{\alpha\beta} W_{\delta\tau} \langle \Psi^\tau | f_r^+ f_h | \Psi^\alpha \rangle W_{\alpha\beta} \langle \Psi^\beta | f_r^+ f_h | \Psi^\delta \rangle$$

So we have,

$$\langle \Psi_0 | A_r^+ A_h | \Psi_0 \rangle \equiv \sum_{\tau\delta} \sum_{\alpha\beta} W_{\delta\tau} \langle \Psi^\tau | M | \Psi^\alpha \rangle W_{\alpha\beta} \langle \Psi^\beta | M^+ | \Psi^\delta \rangle$$

$$= Tr(WMWM^+) \qquad (10)$$

since $\{\Psi_\tau\}$ is a real basis.

By Lieb's theorem, W is a positive definite matrix. Consequently, matrix $W^{1/2}$ is well defined. Therefore we have

$$Tr(WMWM^+) = Tr\left[(W^{1/2}MW^{1/2})(W^{1/2}MW^{1/2})^+\right] \geq 0 \qquad (11)$$

Theorem 1 is proved.

*Remark 1.* It is worthwhile to point out that Theorem 1 has a physical implication. We noticed that

$$B = (B_{rh}) \equiv \left[ \langle \Psi_0(\tfrac{N}{2},\tfrac{N}{2}) | A_r^+ A_h | \Psi_0(\tfrac{N}{2},\tfrac{N}{2}) \rangle \right]$$

is, in fact, the reduced on-site two-particle density matrix of $\tilde{\Psi}_0(\tfrac{N}{2},\tfrac{N}{2})$. If the largest eigenvalues $\lambda_{max}$ of this matrix satisfies the condition $\lambda_{max} \geq \alpha H_\wedge$, where $\alpha$ is a positive constant independent of $H_\wedge$, then $\tilde{\Psi}_0(\tfrac{N}{2},\tfrac{N}{2})$ has an off diagonal long-range order (ODLRO), which indicates that $\tilde{\Psi}_0(\tfrac{N}{2},\tfrac{N}{2})$ is a superfluid. On the other hand, by Theorem 1, we conclude that, if $\tilde{\Psi}_0(\tfrac{N}{2},\tfrac{N}{2})$ is a superfluid, it must be a Bose – Einstein condensate, namely, a macroscopic on-site pair of electrons is condensed at p = 0 in $\tilde{\Psi}_0(\tfrac{N}{2},\tfrac{N}{2})$. A detailed analysis on this point is found in Ref. [16].

Since Theorem 1 hold for any even integer *N*, in particular, it holds for the ground state $\tilde{\Psi}_0(-\varepsilon,-U,\tilde{\Lambda})$ of the negative – ($\varepsilon$,U,) symmetric PAM at half filling. Appling the inverse of the particle-hole transformation (6), we immediately obtain the following theorem.

*Theorem 2:* Let $\Psi_0(\varepsilon,U,\tilde{\Lambda})$ be the ground state of the positive-($\varepsilon,U$) symmetric PAM Hamiltonian at half filling. Define the spin operators of the electron by

$$S_{i+} = f_{i\uparrow}^+ f_{i\downarrow}, S_{i-} = f_{i\downarrow}^+ f_{i\uparrow}, S_{iz} = \frac{1}{2}(f_{i\uparrow}^+ f_{i\uparrow} - f_{i\downarrow}^+ f_{i\downarrow}) \tag{12}$$

$$\langle \Psi_0(\varepsilon, U) | S_{r+} S_{h-} | \Psi_0(\varepsilon, U) \rangle \begin{cases} \geq 0 \text{ if } r \text{ and } h \text{ are in the same sublattice} \\ \leq 0 \text{ if } r \text{ and } h \text{ are in the different sublattice} \end{cases} \tag{13}$$

*Proof.* By the inverse of the unitary particle-hole transformation, we have

$$\Psi_0(-\varepsilon, U, \tilde{\Lambda}) \to \Psi_0(\varepsilon, U, \tilde{\Lambda})$$

$$f_{r\downarrow}^+ f_{r\uparrow}^+ \to \varepsilon(r) f_{r\downarrow} f_{r\uparrow}^+, \quad f_{h\uparrow}^+ f_{h\downarrow} \to \varepsilon(h) f_{h\uparrow}^+ f_{h\downarrow}^+ \tag{14}$$

where $\varepsilon(r) = 1$ if $r \in A$, and $\varepsilon(r) = -1$ if $r \in B$

$$\hat{U}_0 f_{r\uparrow} \hat{U}_0^{-1} = f_{r\uparrow}, \hat{U}_0 f_{r\downarrow} \hat{U}_0^{-1} = \varepsilon(r) f_{r\downarrow}$$

Therefore by Theorem 1,

$$0 \leq \langle \Psi_0(-\varepsilon, -U, \tilde{\Lambda}) | f_{r\downarrow}^+ f_{r\uparrow}^+ f_{h\uparrow} f_{h\downarrow} | \Psi_0(-\varepsilon, -U, \tilde{\Lambda}) \rangle$$

$$= \langle \Psi_0(-\varepsilon, -U, \tilde{\Lambda}) | \hat{U}_0 \hat{U}_0^{-1} (f_{r\downarrow}^+ f_{r\uparrow}^+) \hat{U}_0 \hat{U}_0^{-1} (f_{h\uparrow} f_{h\downarrow}) \hat{U}_0 \hat{U}_0^{-1} | \Psi_0(-\varepsilon, -U, \tilde{\Lambda}) \rangle$$

$$= \varepsilon(r)\varepsilon(h) \langle \Psi_0(\varepsilon, U, \tilde{\Lambda}) | S_{r+} S_{h-} | \Psi_0(\varepsilon, U, \tilde{\Lambda}) \rangle \tag{15}$$

If **r** and **h** belong to the same sublattices, then $\varepsilon(r)\varepsilon(h) = 1$ and hence

$$\langle \Psi_0 | S_{r+} S_{h-} | \Psi_0 \rangle \geq 0$$

Otherwise, $\varepsilon(r)\varepsilon(h) = -1$ and $\langle \Psi_0 | S_{r+} S_{h-} | \Psi_0 \rangle \leq 0$

Theorem 2 is proved.

*Remark 2.* The workers in Ref. [17] used the same technique to show the existence of ferrimagnetism in some positive-U Hubbard model.

Theorem 2 tells us that the short-range transverse spin correlation of *d* or *f* electrons in the ground state of the positive-($\varepsilon$, U,) symmetric PAM is antiferromagnetic. Since the Hamiltonian $\tilde{H}_{\tilde{\Lambda}}$ has the SU(2) spin symmetry and its ground state $\Psi_0(\varepsilon, U, \tilde{\Lambda})$ at half filling is nondegenerate, one would expect that Theorem 2 also holds for the longitudinal spin correlation functions. In fact, we have the following theorem.

*Theorem 3.* Let $\Psi_0(\varepsilon, U, \tilde{\Lambda})$ be the ground state of the positive-($\varepsilon$, U,) symmetric PAM Hamiltonian at half filling. Let

$$\lim_{\varepsilon \to 0} \langle \Psi_0(\varepsilon, U, \tilde{\Lambda}) | S_{rz} S_{hz} | \Psi_0(\varepsilon, U, \tilde{\Lambda}) \rangle \equiv \langle \Psi_0(\varepsilon, U, \tilde{\Lambda}) | S_{rz} S_{hz} | \Psi_0(\varepsilon, U, \tilde{\Lambda}) \rangle \equiv C(r, h); \quad (16)$$

then the spin correlation function C (**r**, **h**) has satisfy inequality (13).

*Proof.* We first show that for any pair of distinct lattice points **r** and **h**,

$$\langle \Psi_0(\varepsilon, U, \tilde{\Lambda}) | S_{r+} S_{h-} | \Psi_0(\varepsilon, U, \tilde{\Lambda}) \rangle = 2 \langle \Psi_0(\varepsilon, U, \tilde{\Lambda}) | S_{rz} S_{hz} | \Psi_0(\varepsilon, U, \tilde{\Lambda}) \rangle \quad (17)$$

holds.

By definition,

$$S_{r+} = S_{rx} + iS_{ry}, S_{r-} = S_{rx} - iS_{ry}, S_{h-} = S_{hx} - iS_{hy}.$$

Therefore we have

$$\langle \Psi_0(\varepsilon, U, \tilde{\Lambda}) | S_{r+} S_{h-} | \Psi_0(\varepsilon, U, \tilde{\Lambda}) \rangle = \langle \Psi_0(\varepsilon, U, \tilde{\Lambda}) | (S_{rx} + iS_{ry})(S_{hx} - iS_{hy}) | \Psi_0(\varepsilon, U, \tilde{\Lambda}) \rangle$$

$$= \langle \Psi_0(\varepsilon, U, \tilde{\Lambda}) | (S_{rx} S_{hx} - iS_{rx} S_{hy} + iS_{ry} S_{hx} + S_{ry} S_{hy}) | \Psi_0(\varepsilon, U, \tilde{\Lambda}) \rangle$$

$$= \langle \Psi_0(\varepsilon, U, \tilde{\Lambda}) | (S_{rx} S_{hx} + S_{ry} S_{hy} + iS_{ry} S_{hx} - iS_{rx} S_{hy}) | \Psi_0(\varepsilon, U, \tilde{\Lambda}) \rangle$$

$$= \langle \Psi_0(\varepsilon, U, \tilde{\Lambda}) | S_{rx} S_{hx} + S_{ry} S_{hy} | \Psi_0(\varepsilon, U, \tilde{\Lambda}) \rangle + $$
$$i \langle \Psi_0(\varepsilon, U, \tilde{\Lambda}) | S_{ry} S_{hx} - S_{rx} S_{hy} | \Psi_0(\varepsilon, U, \tilde{\Lambda}) \rangle \quad (18)$$

We first simplify the last term on the right-hand side of (18). Since **r** and **h** are distinct, [ $S_{rz}$, $S_{hy}$] = [ $S_{ry}$, $S_{hz}$] = 0. Therefore $S_{ry}S_{hx}$ - $S_{rx}$ $S_{hy}$ is a Hermitian operator. Consequently, its expectation value in any state is a real quantity. On the other hand, since $\tilde{H}_{\tilde{\Lambda}}$ is a real matrix, its ground state $\Psi_0(\varepsilon, U, \tilde{\Lambda})$ must be chosen as state real vector. Therefore the expectation value $F$ of $S_{ry}S_{hx}$ - $S_{rx}$ $S_{hy}$ in $\Psi_0(\varepsilon, U, \tilde{\Lambda})$ must be a pure imaginary matrix. Consequently $F \equiv 0$.

Next, we apply the unitary operator

$$\hat{U}_2 = \exp\left[\left(\frac{i\pi}{2}\right)\sum_{i\in\tilde{\lambda}} S_{is}\right]$$

which rotates each spin about the $S_z$ axis by an angle $\pi/2$, to the expectation value of $S_{ry} S_{hy}$ in $\Psi_0(\varepsilon, U, \tilde{\Lambda})$ and obtain

$$\langle\Psi_0(\varepsilon,U,\tilde{\lambda})|S_{ry}S_{hy}|\Psi_0(\varepsilon,U,\tilde{\lambda})\rangle = \langle\Psi_0(\varepsilon,U,\tilde{\lambda})|S_{rx}S_{hx}|\Psi_0(\varepsilon,U,\tilde{\lambda})\rangle \quad (19)$$

Substituting Eq.(19) into Eq.(18), we obtain the identity in Eq.(17), that is:

$$\langle\Psi_0(\varepsilon,U,\tilde{\lambda})|S_{r+}S_{h-}|\Psi_0(\varepsilon,U,\tilde{\lambda})\rangle$$

$$= \langle\Psi_0(\varepsilon,U,\tilde{\lambda})|S_{rx}S_{hx}+S_{ry}S_{hy}|\Psi_0(\varepsilon,U,\tilde{\lambda})\rangle + i\langle\Psi_0(\varepsilon,U,\tilde{\lambda})|S_{ry}S_{hx}-S_{rx}S_{hy}|\Psi_0(\varepsilon,U,\tilde{\lambda})\rangle$$

So we have,

$$\langle\Psi_0(\varepsilon,U,\tilde{\lambda})|S_{r+}S_{h-}|\Psi_0(\varepsilon,U,\tilde{\lambda})\rangle = \langle\Psi_0(\varepsilon,U,\tilde{\lambda})|S_{rx}S_{hx}+S_{ry}S_{hy}|\Psi_0(\varepsilon,U,\tilde{\lambda})\rangle$$

$$= \langle\Psi_0(\varepsilon,U,\tilde{\lambda})|S_{rx}S_{hx}|\Psi_0(\varepsilon,U,\tilde{\lambda})\rangle + \langle\Psi_0(\varepsilon,U,\tilde{\lambda})|S_{ry}S_{hy}|\Psi_0(\varepsilon,U,\tilde{\lambda})\rangle$$

$$= \langle\Psi_0(\varepsilon,U,\tilde{\lambda})|S_{rx}S_{hx}|\Psi_0(\varepsilon,U,\tilde{\lambda})\rangle + \langle\Psi_0(\varepsilon,U,\tilde{\lambda})|S_{rx}S_{hx}|\Psi_0(\varepsilon,U,\tilde{\lambda})\rangle$$

$$= 2\langle\Psi_0(\varepsilon,U,\tilde{\lambda})|S_{rx}S_{hx}|\Psi_0(\varepsilon,U,\tilde{\lambda})\rangle$$

Eq.(17) is proved.

Similarly, we can show that

$$\langle\Psi_0(\varepsilon,U,\tilde{\lambda})|S_{rx}S_{hx}|\Psi_0(\varepsilon,U,\tilde{\lambda})\rangle = \langle\Psi_0(\varepsilon,U,\tilde{\lambda})|S_{rz}S_{hz}|\Psi_0(\varepsilon,U,\tilde{\lambda})\rangle \quad (20)$$

By applying the unitary operator

$$\hat{U}_3 = \exp\left[\left(\frac{i\pi}{2}\right)\sum_{i\in\tilde{\lambda}} S_{iy}\right]$$

To the expectation value of $S_{rz}S_{hz}$ in $\Psi_0(\varepsilon, U, \tilde{\Lambda})$

Combining the identities in Eq.(17) and Eq.(20) and the inequality Eq.(13), we see that the longitudinal spin correlation of $\Psi_0(\varepsilon, U, \tilde{\Lambda})$ is antiferromagnetic. Since this conclusion is true for any $\varepsilon > 0$, it must also hold for the limit $\varepsilon \to 0$. Therefore the longitudinal spin correlation of the *d* or *f* electrons in the ground state of the positive-U symmetric PAM at half filling is antiferromagnetic. Our proof is accomplished.

### 3.0    SUMMARY AND CONCLUSION

This work is an extension and detailed analysis of Tian article [16] and Ueda et. al. [15] for the Hubbard model and PAM respectively. We proved the existence of the short – range antiferromagnetic order in the ground state of the symmetric PAM at half filling for arbitrary V and U > 0. We do not claim that the ground state has long – range antiferromagnetic order. In fact it has been shown, by using a mean field slave – boson theory that the half – filled, that the ground state may be either antiferromagnetic or paramagnetic [18]. The line separating their region of stability is given by the critical value of their exchange constant. In all the lattice systems studied recently, it was observed that they have an antiferromagnetic ground state and the first excited state is always a spin singlet [19].

In conclusion, in this paper, we showed rigorously that the spin correlation between (*d, d*), (*f, f*), (*C, C*), (*C, d*) and (*C, f*) electrons are antiferromagnetic in the ground state of the PAM at half – fillings. The spin correlations of *f* electrons in the ground state $\Psi_0(\wedge)$ which is of fundamental importance was not discussed in this article, which hopefully will become possible in the near future.

**Acknowledgement**

We appreciate the very useful discussions with J. O. A. Idiodi, B. Iyorzor and S. I. Okunzuwa. This work is partially funded by Kusmus Communications, ICBR and AFAHOSITECH.